\documentclass{article}
\usepackage{spconf,amsmath,graphicx,hyperref}
\usepackage{textcomp}
\usepackage{xcolor}
\usepackage{array}
\usepackage{booktabs}
\usepackage{multirow}
\usepackage{tabularx}
\usepackage{makecell}
\usepackage{bm}
\usepackage{amssymb}

\usepackage[table]{xcolor}
\usepackage[dvipsnames]{xcolor}

\title{CONDITION-INVARIANT FMRI DECODING OF SPEECH INTELLIGIBILITY\\ WITH DEEP STATE SPACE MODEL}
\name{Ching-Chih Sung$^{1,3*}$, Shuntaro Suzuki$^{2*}$, Francis Pingfan Chien$^{1,4}$\sthanks{Equal contribution; This work was partially supported by JSPS KAKENHI Grant Number 23K28168 and JST Moonshot.}, Komei Sugiura$^{2}$, Yu Tsao$^{1}$}
\address{{\normalsize\textsuperscript{1}Academia Sinica, Taiwan\quad\textsuperscript{2}Keio University, Japan}\\
{\normalsize\textsuperscript{3}Graduate Institute of Communication
Engineering, National Taiwan University, Taiwan}\\
{\normalsize\textsuperscript{4}Taiwan International Graduate Program in Interdisciplinary Neuroscience, National Taiwan University, Taiwan}}

\begin{document}
%
\maketitle
\begin{abstract}


Clarifying the neural basis of speech intelligibility is critical for computational neuroscience and digital speech processing. 
Recent neuroimaging studies have shown that intelligibility modulates cortical activity beyond simple acoustics, primarily in the superior temporal and inferior frontal gyri.
However, previous studies have been largely confined to clean speech, leaving it unclear whether the brain employs condition-invariant neural codes across diverse listening environments. 
To address this gap, we propose a novel architecture built upon a deep state space model for decoding intelligibility from fMRI signals, specifically tailored to their high-dimensional temporal structure.
We present the first attempt to decode intelligibility across acoustically distinct conditions, showing our method significantly outperforms classical approaches.
Furthermore, region-wise analysis highlights contributions from auditory, frontal, and parietal regions, and cross-condition transfer indicates the presence of condition-invariant neural 
codes, thereby advancing understanding of abstract linguistic representations in the brain.

\end{abstract}
%


\section{Introduction}
 Understanding how the brain decodes speech intelligibility is a fundamental challenge in computational neuroscience and digital speech processing. 
The brain processes speech through a cortical hierarchy, where acoustic features are transformed into abstract linguistic meaning across temporal and fronto-parietal networks~\cite{hickok2007,friederici2011, friederici2020}. 
While neuroimaging studies~\cite{li2023, fedorenko2024} consistently show that activity in these regions, such as the superior temporal gyrus and inferior frontal gyrus, is modulated by intelligibility beyond simple acoustics, most decoding studies have focused on clean speech. 
This leaves a critical question unresolved: does a condition-invariant neural code for intelligibility exist within this network that generalizes across acoustically distinct conditions, such as noisy and enhanced speech? 
Answering this is key to understanding the brain's abstract representation of linguistic meaning.

\begin{figure*}
    \centering
     \includegraphics[width=0.8\linewidth]{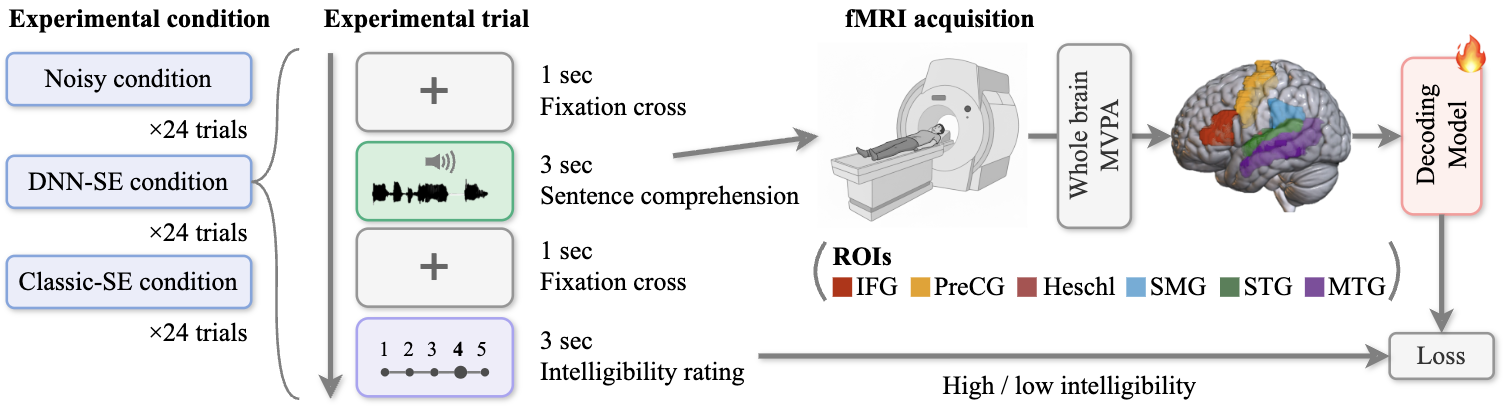}
    \caption{
    Pipeline for fMRI-based decoding of speech intelligibility across acoustically distinct conditions.
    }
    \label{fig:pipeline}
\end{figure*}

From a computational perspective, fMRI-based brain state decoding has traditionally relied on multivoxel pattern analysis (MVPA), which leverages distributed BOLD patterns to infer perceptual and linguistic states~\cite{norman2006}. 
Single-trial generalized linear models further enhance event-related MVPA sensitivity~\cite{lemm2011}, often in combination with robust linear classifiers such as Support Vector Machines (SVMs)~\cite{mumford2012}.
While deep neural networks (DNNs) are well established across many domains and have shown increasing promise in brain decoding~\cite{koyamada2015,liang2024}, their consistent advantage over linear models for fMRI-based intelligibility classification remains unexplored. Moreover, it is unclear whether DNNs capture acoustically condition-invariant neural codes, or which region of interest (ROI) contribute to intelligibility decoding.

In this study, we propose a novel architecture for decoding speech intelligibility from fMRI. 
The architecture is carefully designed to model the high dimensionality of fMRI data and extends deep state space models (deep SSMs)~\cite{S5, Mamba, CausalMamba}, a recent architecture capable of long-sequence modeling. Our main contributions are as follows: 
\begin{itemize}
    \item We present the first attempt to decode speech intelligibility from fMRI across distinct acoustic conditions. 
    \item We introduce a novel architecture based on recent deep SSMs, tailored for intelligibility decoding from fMRI.
    \item Our method consistently outperforms existing approaches ROI-wise and highlights contributions from auditory, frontal, and parietal regions.
    \item We demonstrate cross-condition transfer, implying that our method decodes condition-invariant neural codes.
\end{itemize}

\begin{figure}
    \centering
    \includegraphics[width=0.9\linewidth]{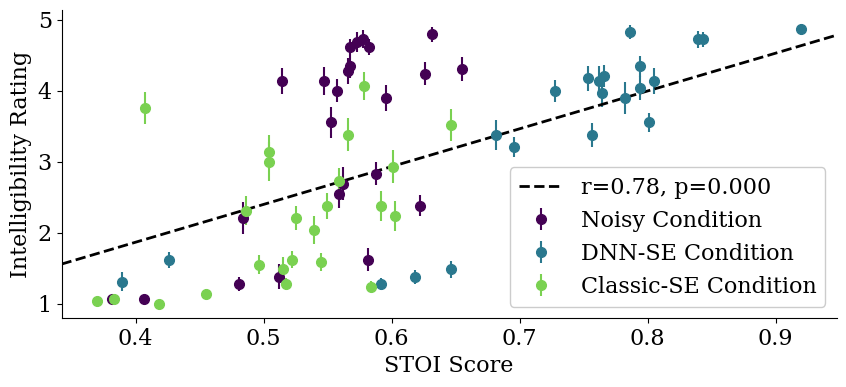}
    \caption{
    Correlation between STOI~\cite{stoi} and perceived speech intelligibility across conditions.
    }
    \label{fig:stoi}
\end{figure}

\section{Experiments}

\subsection{Experimental setup and task design}
We applied our proposed method to an fMRI dataset collected from 25 healthy native Mandarin speakers with normal hearing. 
An overview of the experimental pipeline is shown in Fig.\ref{fig:pipeline}.
During the experiment, participants listened to 72 sentences (10 words each)~\cite{chen2021}, presented under three acoustically distinct conditions: 24 noisy trials (hereafter Noisy condition), 24 deep learning–based speech-enhanced trials (hereafter DNN-SE condition), and 24 classical speech-enhanced trials (hereafter Classic-SE condition).
Stimuli for the Noisy condition were generated by mixing each sentence with stationary speech-shaped noise at –3 dB SNR~\cite{mihai2021}. 
For DNN-SE and Classic-SE conditions, we used SEMamba~\cite{Rong2024}, and the MMSE algorithm~\cite{Ephraim1985}, respectively. 
Stimuli were presented through MRI-compatible headphones.
After each trial, participants rated speech intelligibility on a 5-point scale. Ratings were subsequently binarized (high vs. low) within subject and condition for decoding analyses.

\subsection{fMRI acquisition and pre-processing}
fMRI data were acquired on a Siemens Magnetom Skyra 3T scanner. 
High-resolution anatomical images were collected using a T1-weighted multi-echo magnetization-prepared rapid acquisition gradient echo sequence ($1 \textrm{mm}^{3}$ isotropic). 
Functional scans were acquired with a gradient-echo echo planar imaging sequence (repetition time = 2000 ms, echo time = 24 ms, flip = 90°, field of view = $220 \times 220\ \textrm{mm}^{2}$, 38 slices, voxel size = $3.4 \times 3.4 \times 4.0\ \textrm{mm}^{3}$). 
Pre-processing was conducted in SPM12~\cite{spm12}, including slice-timing correction, motion correction, co-registration to structural images, normalization to Montreal Neurological Institute space, and smoothing with an 8 mm full width at half maximum Gaussian kernel for univariate checks. 
For decoding analyses, unsmoothed single-trial beta maps were extracted using GLMs with trial-wise boxcar regressors convolved with the canonical hemodynamic response function. 
Six motion parameters were included as nuisance regressors.

We focused on 12 bilateral ROIs implicated in speech comprehension~\cite{fedorenko2024}: Heschl’s gyrus (HG), superior temporal gyrus (STG), middle temporal gyrus (MTG), inferior frontal gyrus (IFG), precentral gyrus (PreCG), and supramarginal gyrus (SMG). 
ROI Masks were defined using the Automated Anatomical Labeling atlas and extracted with MarsBaR~\cite{matthew2002}. 

For evaluation, we used four-fold cross-validation and 
models were trained ROI-wise.

\begin{figure}
    \centering
    \includegraphics[width=0.9\linewidth]{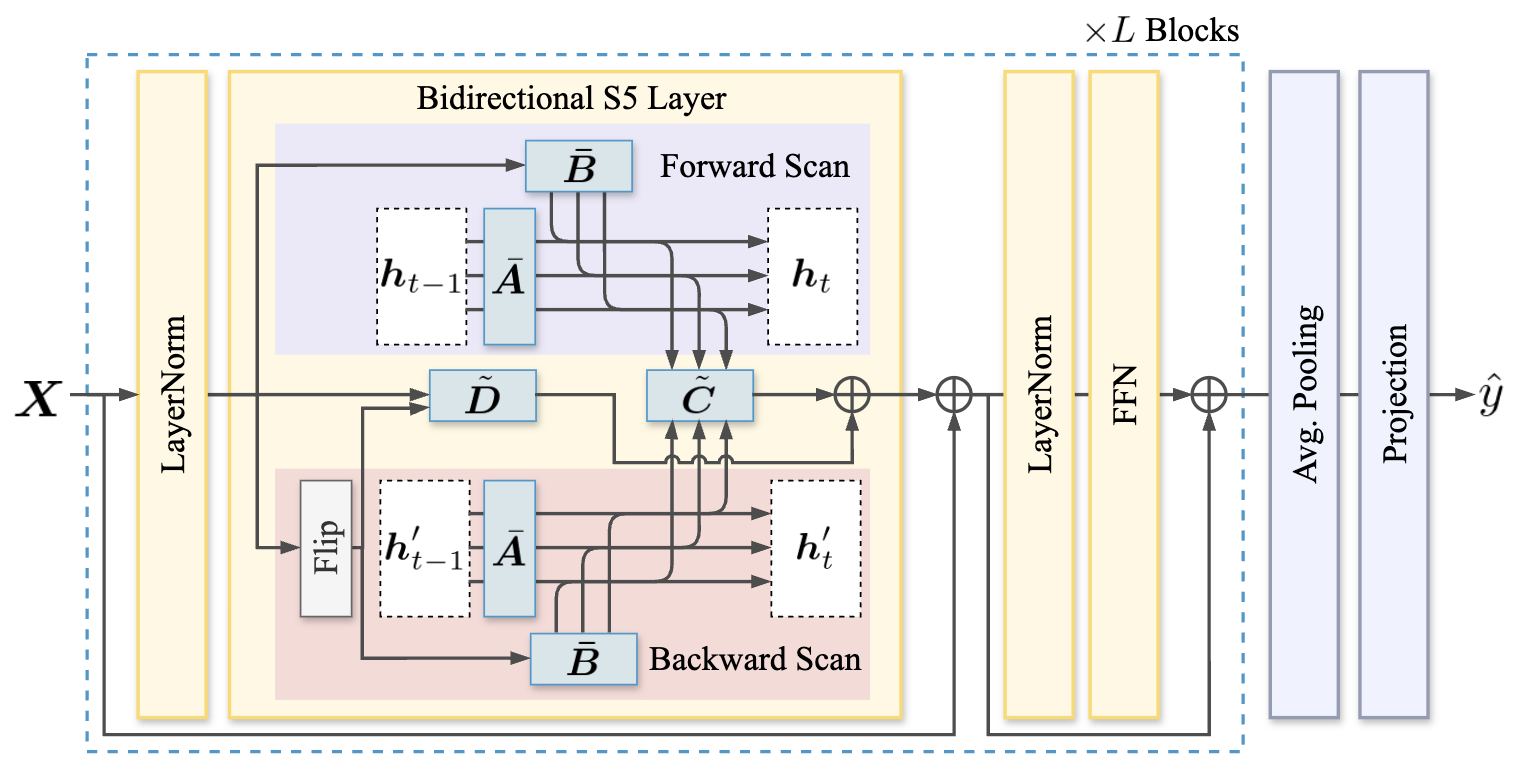}
    \caption{
    Overview of the proposed architecture.
    }
    \label{fig:model_arch}
\end{figure}

\setlength{\tabcolsep}{2.5pt}
\begin{table*}[t]
    \centering
    \caption{Quantitative comparison of decoding performance. \textbf{Bold} values indicate the best performance, and values in brackets denote standard error of the mean. \
    $\dagger$: $p < 0.05$ (ours vs. random);\
    $\ddagger$: $p < 0.05$ (ours vs. best baseline).}
    \label{tab:main}
    \renewcommand{\arraystretch}{0.85}
    \begin{tabular}{l |c c c c c c | c c c c c c }
        \toprule
        \multirow{3}{*}{\textbf{Methods}} & 
        \multicolumn{12}{c}{\textbf{Classification Accuracy [\%]~$\uparrow$}}\\
        &\multicolumn{6}{c|}{Left Hemisphere} 
        &\multicolumn{6}{c}{Right Hemisphere}\\
        & Heschl& STG& MTG& IFG& PreCG& SMG& 
        Heschl& STG& MTG& IFG& PreCG& SMG\\
        \hline
        Random& 50.00& 50.00& 50.00& 50.00& 50.00& 50.00& 50.00& 50.00& 50.00& 50.00& 50.00& 50.00\\
        \hline

        \rowcolor{gray!10}
        \multicolumn{13}{l}{\textit{\color{black!70}Noisy Condition}} \\
        SVM &
        \makecell{57.64 \\(2.48)}&
        \makecell{66.35 \\(3.29)}&
        \makecell{67.40 \\(2.25)}&
        \makecell{59.67 \\(3.06)}&
        \makecell{\textbf{65.21} \\(2.90)}&
        \makecell{62.52 \\(2.56)}&
        \makecell{\textbf{59.50} \\(1.78)}&
        \makecell{67.97 \\(2.66)}&
        \makecell{62.79 \\(2.84)}&
        \makecell{59.23 \\(2.67)}&
        \makecell{70.74 \\(2.90)}&
        \makecell{\textbf{64.27} \\(2.02)}\\
        Transformer &
        \makecell{56.67 \\(2.11)}&
        \makecell{56.67 \\(2.00)}&
        \makecell{60.17 \\(1.72)}&
        \makecell{57.50 \\(2.07)}&
        \makecell{58.67 \\(2.07)}&
        \makecell{58.00 \\(1.34)}&
        \makecell{58.33 \\(1.77)}&
        \makecell{60.33 \\(1.75)}&
        \makecell{56.83 \\(1.88)}&
        \makecell{57.33 \\(2.06)}&
        \makecell{60.17 \\(2.96)}&
        \makecell{58.33 \\(1.85)}\\
        \textbf{Ours} & 
        \makecell{\textbf{58.33}$^\dagger$ \\ (1.98)}&
        \makecell{\textbf{66.50}$^\dagger$ \\(3.19)}&
        \makecell{\textbf{69.83}$^\dagger$ \\(2.78)}&   
        \makecell{\textbf{68.00}$^\dagger$ \\(2.87)}&
        \makecell{63.33$^\dagger$ \\(2.47)}&
        \makecell{\textbf{63.00}$^\dagger$ \\(2.55)}&
        \makecell{\textbf{59.50}$^\dagger$ \\(1.81)}&
        \makecell{\textbf{70.17}$^\dagger$ \\(3.19)}&
        \makecell{\textbf{64.50}$^\dagger$ \\(2.55)}&
        \makecell{\textbf{63.50}$^\dagger$ \\(2.54)}&
        \makecell{\textbf{73.00}$^\dagger$ \\(3.21)}&
        \makecell{62.67$^\dagger$ \\(2.33)}\\

        \hline
        \rowcolor{gray!10}
        \multicolumn{13}{l}{\textit{\color{black!70}DNN-SE Condition}} \\
        SVM &
        \makecell{\textbf{61.70} \\(2.28)}&
        \makecell{64.91 \\(2.72)}&
        \makecell{61.75 \\(2.53)}&
        \makecell{56.95 \\(2.75)}&
        \makecell{61.91 \\(2.39)}&
        \makecell{56.22 \\(2.08)}&
        \makecell{55.63 \\(2.51)}&
        \makecell{\textbf{64.73} \\(3.29)}&
        \makecell{59.83 \\(2.20)}&
        \makecell{59.31 \\(3.05)}&
        \makecell{64.75 \\(2.88)}&
        \makecell{57.28 \\(2.42)}\\
        Transformer &
        \makecell{59.17 \\(2.44)}&
        \makecell{58.33 \\(1.75)}&
        \makecell{52.67 \\(1.56)}&
        \makecell{54.67 \\(1.85)}&
        \makecell{55.33 \\(2.08)}&
        \makecell{53.67 \\(2.00)}&
        \makecell{56.17 \\(2.08)}&
        \makecell{54.00 \\(1.86)}&
        \makecell{58.33 \\(2.19)}&
        \makecell{55.33 \\(1.79)}&
        \makecell{55.17 \\(1.99)}&
        \makecell{54.50 \\(2.03)}\\
        \textbf{Ours} &
        \makecell{57.83$^\dagger$ \\(2.12)}&
        \makecell{\textbf{66.83}$^\dagger$ \\(1.92)}&
        \makecell{\textbf{64.17}$^\dagger$ \\(2.69)}&
        \makecell{\textbf{59.50}$^\dagger$ \\(2.39)}&
        \makecell{\textbf{64.17}$^\dagger$ \\(2.37)}&
        \makecell{\textbf{57.33}$^\dagger$ \\(2.31)}&
        \makecell{\textbf{58.33}$^\dagger$ \\(1.97)}&
        \makecell{64.33$^\dagger$ \\(2.65)}&
        \makecell{\textbf{61.83}$^\dagger$ \\(2.94)}&
        \makecell{\textbf{60.67}$^\dagger$ \\(2.57)}&
        \makecell{\textbf{67.50}$^\dagger$ \\(2.89)}&
        \makecell{\textbf{62.33}$^{\dagger\ddagger}$ \\(2.86)}\\

        \hline
        \rowcolor{gray!10}
        \multicolumn{13}{l}{\textit{\color{black!70}Classic-SE Condition}} \\
        SVM &
        \makecell{57.71 \\(2.73)}&
        \makecell{61.30 \\(2.80)}&
        \makecell{63.47 \\(2.49)}&
        \makecell{59.54 \\(2.45)}&
        \makecell{59.66 \\(2.33)}&
        \makecell{55.64 \\(2.93)}&
        \makecell{56.75 \\(2.44)}&
        \makecell{60.92 \\(3.01)}&
        \makecell{60.51 \\(2.54)}&
        \makecell{58.77 \\(2.41)}&
        \makecell{60.42 \\(2.86)}&
        \makecell{57.21 \\(2.14)}\\
        Transformer &
        \makecell{61.50 \\(2.40)}&
        \makecell{64.50 \\(2.79)}&
        \makecell{60.00 \\(3.38)}&
        \makecell{60.67 \\(2.81)}&
        \makecell{61.17 \\(3.27)}&
        \makecell{57.67 \\(2.88)}&
        \makecell{60.50 \\(2.58)}&
        \makecell{59.17 \\(3.39)}&
        \makecell{63.83 \\(2.91)}&
        \makecell{57.17 \\(2.95)}&
        \makecell{63.83 \\(2.17)}&
        \makecell{64.00 \\(2.90)}\\
        \textbf{Ours} &
        \makecell{\textbf{68.50}$^{\dagger\ddagger}$ \\(2.92)}&
        \makecell{\textbf{67.67}$^\dagger$ \\(2.46)}&
        \makecell{\textbf{66.17}$^\dagger$ \\(3.38)}&
        \makecell{\textbf{64.50}$^\dagger$ \\(2.94)}&
        \makecell{\textbf{64.50}$^\dagger$ \\(2.68)}&
        \makecell{\textbf{64.33}$^\dagger$ \\(2.93)}&
        \makecell{\textbf{65.33}$^\dagger$ \\(2.94)}&
        \makecell{\textbf{70.17}$^{\dagger\ddagger}$ \\(2.38)}&
        \makecell{\textbf{66.50}$^\dagger$ \\(2.38)}&
        \makecell{\textbf{65.50}$^{\dagger\ddagger}$ \\(2.48)}&
        \makecell{\textbf{68.50}$^\dagger$ \\(2.67)}&
        \makecell{\textbf{65.33}$^\dagger$ \\(2.73)}\\
        \bottomrule         
    \end{tabular}
\end{table*}

\subsection{Behavioral and objective validation}
Inside the scanner, participants’ perceived intelligibility closely tracked an objective metric across all three speech conditions. 
For each sentence, in-scanner ratings (1–5) were averaged across 25 subjects and compared with short-time objective intelligibility (STOI)~\cite{stoi}. 
Ratings correlated strongly with STOI across the Noisy, DNN-SE, and Classic-SE conditions ($r = 0.78$, $p < 10^{-4}$; Fig.~\ref{fig:stoi}). 
This tight correspondence demonstrates that subjective ratings provide a reliable index of objective intelligibility, thereby validating the behavioral target for subsequent fMRI decoding.
\section{Proposed Method}
\subsection{Model architecture}
Fig.~\ref{fig:model_arch} illustrates the architecture of our proposed method desgined for speech intelligibility decoding from fMRI. 
It extends S5~\cite{S5}, a variant of deep SSMs~\cite{S4, Mamba}, to effectively model long-range voxel sequences within each ROI of fMRI.

Deep SSMs, inspired by control-theoretic state space formulations~\cite{KalmanFilter}, are parallelizable recurrent neural networks that have shown strong performance in modeling long-range dependencies~\cite{LSSL, S4}. 
The recent success of Mamba~\cite{Mamba, Mamba2}, a deep SSM variant, in language modeling has further accelerated the adoption of this framework across various domains~\cite{SaShiMi, VMamba}.
In our study, voxel counts per ROI are particularly large, reaching the highest voxel count of 11,669 in the MTG, motivating the need for architectures capable of handling long sequences. 
To this end, we extend S5, a representative deep SSM well-suited for modeling continuous signals.

In S5, the input fMRI signals $\mathbf{x}_t \in \mathbb{R}^P$ are mapped to output signals $\mathbf{y}_t \in \mathbb{R}^P$ via latent states $\mathbf{h}_t \in \mathbb{R}^Q$ as follows:
\begin{align}
    \bm{h}_t = \bar{\bm{A}}\bm{h}_{t-1} + \bar{\bm{B}}\bm{x}_t,\quad \bm{y}_t = \bar{\bm{C}}\bm{h}_t + \bar{\bm{D}}\bm{x}_t.
\end{align}
Here, $P$ and $Q$ represent the dimensions of the input/output signals and the latent states, respectively, with $Q=rP$, where $r$ denotes the state expansion ratio.
The matrices $\bar{\mathbf{A}} \in \mathbb{R}^{Q \times Q}$, $\bar{\mathbf{B}} \in \mathbb{R}^{Q \times P}$, $\bar{\mathbf{C}} \in \mathbb{R}^{P \times Q}$, and $\bar{\mathbf{D}} \in \mathbb{R}^{P \times P}$ represent the discretized state transition matrices.
However, as fMRI consist of voxel sequences and lack unidirectional causality, we extend the above framework to a bidirectional formulation (here after denoted as S5 Bidir.) as follows:
\begin{equation}
\begin{aligned}
    \bm{h}_t   &= \bar{\bm{A}}\bm{h}_{t-1} + \bar{\bm{B}}\bm{x}_t,\quad
    \bm{h}_t'  = \bar{\bm{A}}\bm{h}_{t-1}' + \bar{\bm{B}}\bm{x}_{T-t}, \\
    \bm{y}_t   &= \tilde{\bm{C}}[\bm{h}_t;\bm{h}_t'] + \tilde{\bm{D}}[\bm{x}_t;\bm{x}_{T-t}].
\end{aligned}
\end{equation}
Here, $\mathbf{h}_t'\in\mathbb{R}^Q$ denotes the latent state for the backward scan, and $\tilde{\mathbf{C}} \in \mathbb{R}^{P \times 2Q}$ and $\tilde{\mathbf{D}} \in \mathbb{R}^{P \times 2P}$ represent the modulated state transition matrices.

Building on the above, our proposed method employs a stack of $L$ blocks, each comprising LayerNorm~\cite{LayerNorm}, S5 Bidir., and a feed-forward network.
The resulting representations are subsequently processed by average pooling and a projection layer to predict the probability $\hat{y}$ that the subject’s intelligibility was high at the time of fMRI acquisition. 

\subsection{Implementation details}
We employed the AdamW optimizer~\cite{adamw} with a learning rate of $5.0\times 10^{-6}$. 
The batch size was set to 8 and the training ran for 50 epochs. 
We set the stacked block number $L=2$ and the hidden state expansion ratio $r=2.0$.

\begin{figure}
    \centering
    \includegraphics[width=0.9\linewidth]{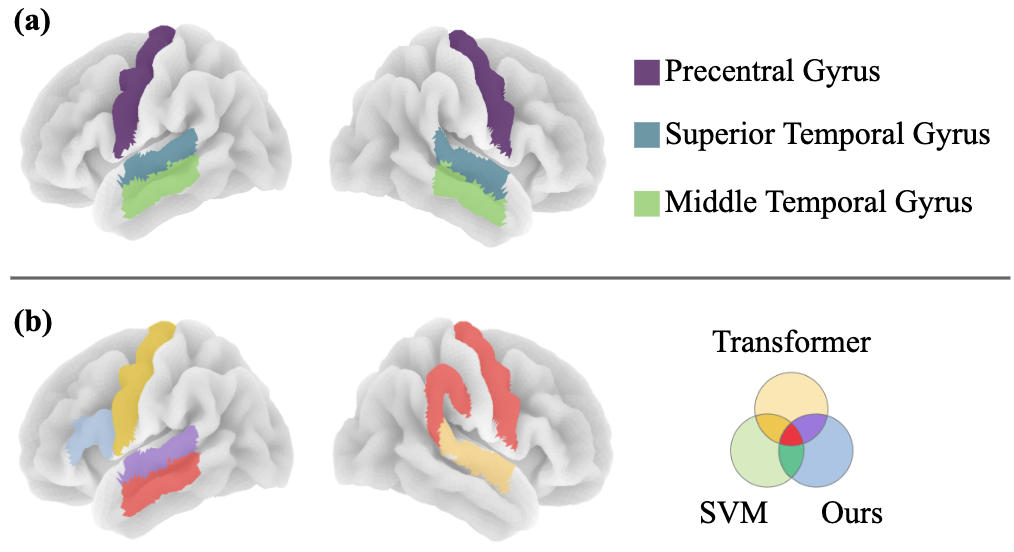}
    \caption{
    Visualization of significant ROIs in speech intelligibility decoding. (a) Whole-brain MVPA results (family-wise error corrected, $p < 0.001$). (b) Top five ROIs with the highest decoding performance in the Noisy condition.
    }
    \label{fig:mvpa}
\end{figure}

\section{Results and discussion}

\subsection{Within-condition decoding}

Table~\ref{tab:main} presents the ROI-wise comparison of decoding performance between our proposed method and two established baselines: an L2-regularized linear SVM~\cite{SVM} and a Transformer~\cite{transformer}. Across the 12 bilateral ROIs, the proposed method was consistently competitive, achieving the highest accuracy in 10, 10, and 12 ROIs for the Noisy, DNN-SE, and Classic-SE conditions, respectively. In the Noisy condition, the strongest decoding was observed in the right hemisphere, with STG reaching 70.17\% and PreCG reaching 73.00\%. All results were significantly above chance ($p < 0.05$) and typically exceeded the best baseline in pairwise comparisons within ROI ($p < 0.05$).
Moreover, these ROIs consistently showed significant performance gains in the DNN-SE and Classic-SE conditions, indicating that enhanced speech continues to engage distributed cortical patterns underlying perceived intelligibility, and that our method effectively exploits these patterns. 
Together, these results within conditions highlight a distributed cortical substrate for intelligibility decoding that spans the temporal cortex and extends into the frontal and parietal regions. 

Furthermore, Fig.~\ref{fig:mvpa} shows the results of MVPA analyses of the whole brain, revealing higher decoding performance in bilateral STG, MTG, and PreCG, as well as consistent top performance in all three models in the Noisy condition.

\setlength{\tabcolsep}{2.5pt}
\begin{table}[t]
    \centering
    \caption{Cross-condtion transfer results on left hemisphere. \
    $\dagger$: $p < 0.05$ (ours vs. random).}
    \label{tab:transfer}
    \renewcommand{\arraystretch}{0.85}
    \begin{tabular}{l |c c c c c c}
        \toprule
        \multirow{2}{*}{\textbf{Methods}} & 
        \multicolumn{6}{c}{\textbf{Classification Accuracy [\%]~$\uparrow$}}\\
        & Heschl& STG& MTG& IFG& PreCG& SMG\\
        \hline
        Random& 50.00& 50.00& 50.00& 50.00& 50.00& 50.00\\
        \hline
        \makecell[l]{Noisy$\rightarrow$\\DNN-SE}&
        \makecell{58.63$^\dagger$ \\(2.09)}&
        \makecell{60.25$^\dagger$ \\(2.08)}&
        \makecell{61.42$^\dagger$ \\(2.08)}&
        \makecell{59.25$^\dagger$ \\(1.90)}&
        \makecell{61.58$^\dagger$ \\(2.25)}&
        \makecell{57.08$^\dagger$ \\(2.26)}\\
        \makecell[l]{Noisy$\rightarrow$\\Classic-SE}&
        \makecell{51.13 \\(2.86)}&
        \makecell{61.71$^\dagger$ \\(1.95)}&
        \makecell{61.54$^\dagger$ \\(2.50)}&
        \makecell{58.67$^\dagger$ \\(2.61)}&
        \makecell{59.67$^\dagger$ \\(1.80)}&
        \makecell{54.88$^\dagger$ \\(2.22)}\\
        \bottomrule
    \end{tabular}
\end{table}
\setlength{\tabcolsep}{2.5pt}
\begin{table}[t]
    \centering
    \caption{Ablation study of the proposed method on the left hemisphere under the Noisy condition.
    Bidir. denotes bidirectional scanning. 
    \ $\dagger$: $p < 0.05$ (ours vs. random).}
    \label{tab:ablation}
    \renewcommand{\arraystretch}{0.85}
    \begin{tabular}{l |c c c c c c}
        \toprule
        \multirow{2}{*}{\textbf{Methods}} & 
        \multicolumn{6}{c}{\textbf{Classification Accuracy [\%]~$\uparrow$}}\\
        & Heschl& STG& MTG& IFG& PreCG& SMG\\
        \hline
        Random& 50.00& 50.00& 50.00& 50.00& 50.00& 50.00\\
        \hline
        \makecell[l]{Ours}&
        \makecell{\textbf{58.33} \\(1.98)}&
        \makecell{\textbf{66.50} \\(3.19)}&
        \makecell{69.83 \\(2.78)}&
        \makecell{\textbf{68.00} \\(2.87)}&
        \makecell{63.33 \\(2.47)}&
        \makecell{\textbf{63.00} \\(2.55)}\\
        \makecell[l]{w/o Bidir.}&
        \makecell{57.17 \\(2.03)}&
        \makecell{65.67 \\(3.04)}&
        \makecell{70.33 \\(2.57)}&
        \makecell{65.83 \\(3.13)}&
        \makecell{\textbf{65.33} \\(2.32)}&
        \makecell{62.67 \\(2.38)}\\
        \makecell[l]{w/o S5}&
        \makecell{57.33 \\(2.00)}&
        \makecell{65.83 \\(3.05)}&
        \makecell{\textbf{70.67} \\(2.64)}&
        \makecell{65.50 \\(3.11)}&
        \makecell{65.00 \\(2.39)}&
        \makecell{62.67 \\(2.38)}\\
        \bottomrule 
    \end{tabular}
\end{table}
\subsection{Cross-condition transfer decoding}
We next tested whether the proposed method trained on the Noisy condition generalizes to enhanced speech conditions (Table~\ref{tab:transfer}). 
Transfer was significantly above chance ($p<0.05$) across multiple ROIs for both the Noisy to DNN-SE condition and the Noisy to Classic-SE condition. 
For the Noisy to DNN-SE condition, PreCG reached the highest performance at 61.58\%. 
For the Noisy to Classic-SE condition, STG reached the highest performance of 61.71\%. 
These results show that the discriminative neural code exploited by the proposed method is not tied to the acoustic profile of a specific enhancement algorithm, but instead resides in higher-level representations within STG, MTG, IFG, and PreCG.

\subsection{Ablation Study}
Table~\ref{tab:ablation} summarizes the ablation study on the left hemisphere under the Noisy condition, comparing variants without (i) bidirectional scanning and (ii) the S5 layer. The full model achieved the highest or tied accuracy in four of six ROIs (HG, STG, IFG, SMG). These results suggest that both bidirectionality and the S5 layer contribute effectively to decoding speech intelligibility from fMRI.
\section{Conclusion}

In this study, we addressed the decoding of speech intelligibility from fMRI signals under noisy and enhanced speech conditions. 
Our method consistently outperformed baselines, with the largest gains in the temporal cortex and precentral gyrus. 
Notably, models trained on noisy speech generalized to enhanced speech, suggesting a condition-invariant neural code. 
These findings highlight the potential of brain-informed tuning, where neural decoding can guide speech enhancement to improve intelligibility under degraded inputs. Future work should integrate EEG/MEG for higher temporal resolution and embed brain-derived signals into real-time enhancement.

\vfill\pagebreak
\bibliographystyle{IEEEbib}
\bibliography{refs}

\begin{thebibliography}{10}

\bibitem{hickok2007}
Gregory Hickok et~al.,
\newblock ``{The cortical organization of speech processing},''
\newblock {\em Nat. Rev. Neurosci.}, vol. 8, no. 5, pp. 393--402, 2007.

\bibitem{friederici2011}
Angela~D Friederici,
\newblock ``{The Brain Basis of Language Processing: From Structure to Function},''
\newblock {\em Physiol. Rev.}, vol. 91, no. 4, pp. 1357--1392, 2011.

\bibitem{friederici2020}
Angela~D Friederici,
\newblock ``{Hierarchy processing in human neurobiology: how specific is it?},''
\newblock {\em Phil. Trans. R. Soc. Lond. B Biol. Sci.}, vol. 375, no. 1789, pp. 20180391, 2020.

\bibitem{li2023}
Yuanning Li et~al.,
\newblock ``{Dissecting neural computations in the human auditory pathway using deep neural networks for speech},''
\newblock {\em Nat. Neurosci.}, vol. 26, no. 12, pp. 2213--2225, 2023.

\bibitem{fedorenko2024}
Evelina Fedorenko et~al.,
\newblock ``{The language network as a natural kind within the broader landscape of the human brain},''
\newblock {\em Nat. Rev. Neurosci.}, vol. 25, no. 5, pp. 289--312, 2024.

\bibitem{norman2006}
Kenneth~A Norman et~al.,
\newblock ``{Beyond mind-reading: multi-voxel pattern analysis of fMRI data},''
\newblock {\em Trends Cogn. Sci.}, vol. 10, no. 9, pp. 424--430, 2006.

\bibitem{lemm2011}
Steven Lemm et~al.,
\newblock ``{Introduction to machine learning for brain imaging},''
\newblock {\em NeuroImage}, vol. 56, no. 2, pp. 387--399, 2011.

\bibitem{mumford2012}
Jeanette~A. Mumford et~al.,
\newblock ``{Deconvolving BOLD activation in event-related designs for multivoxel pattern classification analyses},''
\newblock {\em NeuroImage}, vol. 59, no. 3, pp. 2636--2643, 2012.

\bibitem{koyamada2015}
Sotetsu Koyamada et~al.,
\newblock ``{Deep learning of fMRI big data: a novel approach to subject-transfer decoding},''
\newblock {\em arXiv preprint arXiv:1502.00093}, 2015.

\bibitem{liang2024}
Yun Liang et~al.,
\newblock ``{Decoding fMRI data with support vector machines and deep neural networks},''
\newblock {\em J. Neurosci.}, vol. 401, pp. 110004, 2024.

\bibitem{S5}
Jimmy~T.H. Smith et~al.,
\newblock ``{Simplified State Space Layers for Sequence Modeling},''
\newblock in {\em ICLR}, 2023.

\bibitem{Mamba}
Albert Gu et~al.,
\newblock ``{Mamba: Linear-Time Sequence Modeling with Selective State Spaces},''
\newblock in {\em CoLM}, 2024.

\bibitem{CausalMamba}
Weihao Deng et~al.,
\newblock ``{Causal fMRI-Mamba: Causal State Space Model for Neural Decoding and Brain Task States Recognition},''
\newblock in {\em ICASSP}, 2025, pp. 1--5.

\bibitem{stoi}
Cees~H. Taal et~al.,
\newblock ``{A short-time objective intelligibility measure for time-frequency weighted noisy speech},''
\newblock in {\em ICASSP}, 2010, pp. 4214--4217.

\bibitem{chen2021}
Ryandhimas~E. Zezario et~al.,
\newblock ``{Deep Learning-Based Non-Intrusive Multi-Objective Speech Assessment Model With Cross-Domain Features},''
\newblock {\em IEEE/ACM Trans. Audio Speech Lang. Process.}, vol. 31, pp. 54--70, 2023.

\bibitem{mihai2021}
Paul~Glad Mihai et~al.,
\newblock ``{Modulation of the Rrimary Auditory Thalamus When Recognizing Speech With Background Noise},''
\newblock {\em J. Neurosci.}, vol. 41, no. 33, pp. 7136--7147, 2021.

\bibitem{Rong2024}
Rong Chao et~al.,
\newblock ``{An Investigation of Incorporating Mamba For Speech Enhancement},''
\newblock in {\em SLT Workshop}, 2024, pp. 302--308.

\bibitem{Ephraim1985}
Y.~Ephraim et~al.,
\newblock ``{Speech enhancement using a minimum mean-square error log-spectral amplitude estimator},''
\newblock {\em IEEE TASLPRO}, vol. 33, no. 2, pp. 443--445, 1985.

\bibitem{spm12}
Karl~J. Friston et~al.,
\newblock {\em {Statistical Parametric Mapping: The Analysis of Functional Brain Images}},
\newblock Academic Press (Elsevier), 2007.

\bibitem{matthew2002}
Matthew Brett et~al.,
\newblock ``{Region of interest analysis using the MarsBar toolbox for SPM 99},''
\newblock {\em Neuroimage}, vol. 16, 01 2002.

\bibitem{S4}
Albert Gu et~al.,
\newblock ``{Efficiently Modeling Long Sequences with Structured State Spaces},''
\newblock in {\em ICLR}, 2022.

\bibitem{KalmanFilter}
R.~Kalman,
\newblock ``{A New Approach to Linear Filtering and Prediction Problems},''
\newblock {\em J. Basic.}, vol. 82, no. 1, pp. 35--45, 1960.

\bibitem{LSSL}
Albert Gu et~al.,
\newblock ``{Combining Recurrent, Convolutional, and Continuous-time Models with Linear State-Space Layers},''
\newblock in {\em NeurIPS}, 2021, vol.~34, pp. 572--585.

\bibitem{Mamba2}
Tri Dao et~al.,
\newblock ``{Transformers are SSMs: Generalized Models and Efficient Algorithms Through Structured State Space Duality},''
\newblock in {\em ICML}, 2024, vol. 235, pp. 10041--10071.

\bibitem{SaShiMi}
Karan Goel et~al.,
\newblock ``{It’s Raw! Audio Generation with State-Space Models},''
\newblock in {\em ICML}, 2022, pp. 7616--7633.

\bibitem{VMamba}
Yue Liu et~al.,
\newblock ``{VMamba: Visual State Space Model},''
\newblock in {\em NeurIPS}, 2024.

\bibitem{LayerNorm}
Jimmy~Lei Ba,
\newblock ``{Layer normalization},''
\newblock {\em arXiv preprint arXiv:1607.06450}, 2016.

\bibitem{adamw}
Ilya Loshchilov et~al.,
\newblock ``{Decoupled weight decay regularization},''
\newblock in {\em ICLR}, 2019.

\bibitem{SVM}
Corinna Cortes et~al.,
\newblock ``{Support-vector networks},''
\newblock {\em Machine learning}, vol. 20, no. 3, pp. 273--297, 1995.

\bibitem{transformer}
Ashish Vaswani et~al.,
\newblock ``{Attention is all you need},''
\newblock {\em NeurIPS}, vol. 30, 2017.

\end{thebibliography}

\end{document}